\documentclass{sig-alternate}
\usepackage{algorithmic}
\usepackage{algorithm}
\usepackage{verbatim}

\newcommand{\E}{\mathrm{E}}
\newcommand{\e}{\mathcal{E}}

\newcommand{\Vol}{\mathrm{Vol}}
\newcommand{\eps}{\epsilon}
\newcommand{\bR}{\mathbb{R}}

\newdef{definition}{Definition}
\newtheorem{theorem}{Theorem}[section]
\newtheorem{lemma}[theorem]{Lemma}
\newtheorem{proposition}[theorem]{Proposition}
\newtheorem{corollary}[theorem]{Corollary}

\newenvironment{prevproof}[2]{\noindent {\em {Proof of
{#1}~\ref{#2}:}}}{$\blacksquare$\vskip \belowdisplayskip}

\def\vec{}
\def\eps{\epsilon}
\def\sse{\subseteq}
\def\sm{\setminus}

\def\Lap{\textrm{Lap}}
\def\out{(\vec{d},\vec{a})}
\def\outi{(\vec{d}_i,\vec{a}_i}
\def\outi1{(\vec{d}_{i-1},\vec{a}_{i-1}}

\begin{document}
\conferenceinfo{STOC'10,} {June 5--8, 2010, Cambridge, Massachusetts, USA.}
\CopyrightYear{2010}
\crdata{978-1-4503-0050-6/10/06}
\clubpenalty=10000
\widowpenalty = 10000

\numberofauthors{2}
\title{Interactive Privacy via the Median Mechanism}
\author{
\alignauthor
Aaron Roth\titlenote{Supported in part by an NSF Graduate Research Fellowship. Portions of this work were done while visiting Stanford University.}\\
       \affaddr{Carnegie Mellon University} \\
       \affaddr{5000 Forbes Avenue}\\
       \affaddr{Pittsburgh, PA 15217}\\
       \email{alroth@cs.cmu.edu}
\alignauthor
Tim Roughgarden\titlenote{Supported in part by NSF CAREER Award CCF-0448664, an ONR Young Investigator Award, an ONR PECASE Award, an AFOSR MURI grant, and an Alfred P. Sloan Fellowship.}\\
       \affaddr{Stanford University}\\
       \affaddr{353 Serra Mall}\\
       \affaddr{Stanford, CA 94305}\\
       \email{tim@cs.stanford.edu}
}
\maketitle
\sloppy
\begin{abstract}
We define a new interactive differentially private mechanism
--- the {\em median mechanism} --- for answering arbitrary predicate
queries that arrive online.
Given fixed accuracy and privacy constraints, this mechanism
can answer exponentially more queries than the previously best known
interactive privacy mechanism (the Laplace mechanism, which
independently perturbs each query result).
With respect to the number of queries, our guarantee is close to the best possible, even for non-interactive
privacy mechanisms.
Conceptually, the median mechanism is the first privacy mechanism
capable of identifying and exploiting correlations among queries in
an interactive setting.

We also give an efficient implementation of the median mechanism,
with running time polynomial in the number of queries, the database
size, and the domain size.  This efficient implementation guarantees
privacy for
all input databases, and accurate query results for almost all input
distributions.  The dependence of the privacy on the number of queries in
this mechanism improves over that of the best previously known
efficient mechanism by a super-polynomial factor, even in the
non-interactive setting.
\end{abstract}



%\keywords{Differential Privacy}

\section{Introduction}

Managing a data set with sensitive but useful information,
such as medical records, requires reconciling two objectives:
providing utility to others, perhaps in the form of aggregate
statistics; and respecting the privacy of individuals who
contribute to the data set.
The field of {\em private data analysis}, and in particular
work on {\em differential privacy}, provides a mathematical foundation
for reasoning about this utility-privacy trade-off and offers methods
for non-trivial data analysis that are provably privacy-preserving in
a precise sense.
For a recent survey of the field, see Dwork~\cite{DworkSurvey}.

More precisely, consider a domain $X$ and database size $n$.
A {\em mechanism} is a randomized function from the set $X^n$ of
databases to some range.
For a parameter $\alpha > 0$, a mechanism $M$ is {\em
$\alpha$-differentially private} if, for every database $D$ and fixed
subset $S$ of the range of $M$, changing a single component of
$D$ changes the probability that $M$ outputs something in $S$ by at
most an $e^{\alpha}$ factor.
The output of a differentially private mechanism (and any analysis or
privacy attack that follows) is thus essentially independent of
whether or not a given individual ``opts in'' or ``opts out'' of the
database.

Achieving differential privacy requires ``sufficiently noisy''
answers~\cite{DN03}.  For example, suppose we're interested in the
result of a query $f$ --- a function from databases to some
range --- that simply counts the fraction of database elements that
satisfy some predicate $\varphi$ on $X$.  A special case of a result
in Dwork et al.~\cite{DMNS06} asserts that the following mechanism is
$\alpha$-differentially private: if the underlying database is $D$,
output $f(D) + \Delta$, where the output perturbation $\Delta$ is
drawn from the Laplace distribution $\Lap(\tfrac{1}{n\alpha})$ with
density $p(y) = \tfrac{n\alpha}{2} \exp(-n\alpha|y|)$.
Among all $\alpha$-differentially private mechanisms,
this one (or rather, a discretized analog of it) maximizes user
utility in a strong sense~\cite{GRS09}.

What if we care about more than a single one-dimensional statistic?
Suppose we're interested in $k$ predicate queries $f_1,\ldots,f_k$,
where $k$ could be large, even super-polynomial in $n$.
A natural solution is to use an independent Laplace perturbation for
each query answer~\cite{DMNS06}.  To maintain $\alpha$-differential
privacy, the
magnitude of noise has to scale linearly with $k$, with each
perturbation drawn from $\Lap(\tfrac{k}{n\alpha})$.  Put another way,
suppose one fixes ``usefulness parameters'' $\eps,\delta$,
and insists that the mechanism is {\em $(\eps,\delta)$-useful},
meaning that the outputs are within $\eps$ of the
correct query answers with probability at least $1-\delta$.
This constrains the magnitude of the Laplace noise,
and the
privacy parameter $\alpha$ now suffers linearly with the number $k$
of answered queries.
This dependence limits the use of this mechanism to a sublinear
$k=o(n)$ number of queries.

Can we do better than independent output perturbations?
For special classes of queries like predicate queries, Blum, Ligett,
and Roth~\cite{BLR08} give an affirmative answer (building on
techniques of Kasiviswanathan et al.~\cite{KLNRS09}).
Specifically, in~\cite{BLR08} the exponential mechanism of McSherry
and Talwar~\cite{MT07} is used to show that, for fixed usefulness
parameters $\eps,\delta$, the privacy parameter $\alpha$ only has to
scale {\em logarithmically} with the number of queries.\footnote{More
generally, linearly with the VC dimension of the set of queries, which is
always at most $\log_2 k$.}
This permits simultaneous non-trivial utility and privacy guarantees
even for an exponential number of queries.
Moreover, this dependence on $\log k$ is necessary
in every differentially private mechanism
(see the full version of~\cite{BLR08}).

The mechanism in~\cite{BLR08} suffers from two drawbacks, however.
First, it is {\em non-interactive}: it requires all queries
$f_1,\ldots,f_k$ to be given up front, and computes (noisy)
outputs of all of them at once.\footnote{Or rather, it computes a compact
representation of these outputs in the form of a synthetic database.}
By contrast, independent
Laplace output perturbations can obviously be implemented
interactively, with the queries arriving online and each
answered immediately.
There is good intuition for why the non-interactive setting helps:
outperforming independent output perturbations requires
correlating perturbations across multiple queries, and this is clearly
easier when the queries are known in advance.
Indeed, prior to the
present work, no interactive mechanism better than independent Laplace
perturbations was known.

Second, the mechanism in~\cite{BLR08} is {\em inefficient}.
Here by ``efficient'' we mean has running time polynomial in $n$, $k$,
and $|X|$; Dwork et al.~\cite{DNRRV09} prove that this
is essentially the best one could hope for (under certain cryptographic
assumptions).  The mechanism in~\cite{BLR08} is not efficient because
it requires sampling from
a non-trivial probability distribution over an
unstructured space of exponential size.  Dwork et al.~\cite{DNRRV09}
recently gave an efficient (non-interactive) mechanism that is better
than independent Laplace perturbations, in that the privacy parameter
$\alpha$ of the mechanism scales as $2^{\sqrt{\log k}}$ with the
number of queries~$k$ (for fixed usefulness parameters~$\eps,\delta$).

Very recently, Hardt and Talwar \cite{HT10} gave upper and lower bounds for answering noninteractive linear queries which are tight in a related setting. These bounds are not tight in our setting, however, unless the number of queries is small with respect to the size of the database. When the number of queries is large, our mechanism actually yields error significantly less than required in general by their lower bound \footnote{We give a mechanism for answering $k$ ``counting queries'' with coordinate-wise error $O(n^{2/3}\log k(\log |X|/\alpha)^{1/3})$. This is less error than required by their \emph{lower} bound of roughly $\Omega(\sqrt{k\log (|X|/k)}/\alpha)$ unless $k \leq \tilde{O}((n\alpha/\log |X|)^{4/3})$. We can take $k$ to be as large as $k = \tilde{\Omega}(2^{(n\alpha/\log |X|)^{1/3}})$, in which case our upper bound is a significant improvement -- as are the upper bounds of \cite{BLR08} and \cite{DNRRV09}.}. This is not a contradiction, because when translated into the setting of \cite{HT10}, our database size $n$ becomes a sparsity parameter that is not considered in their bounds.
\vfill\eject
\subsection{Our Results}

We define a new interactive differentially private mechanism for
answering $k$ arbitrary
predicate queries, called the {\em median
mechanism}.\footnote{The privacy guarantee is
$(\alpha,\tau)$-differential
privacy for a negligible function~$\tau$; see Section~\ref{sec:prelim}
for definitions.}
The basic implementation of the median mechanism interactively answers
queries $f_1,\ldots,f_k$ that arrive online, is
$(\eps,\delta)$-useful, and has privacy $\alpha$ that scales
with $\log k \log |X|$; see Theorem~\ref{thm:main1} for the exact
statement.
These privacy and utility guarantees hold
even if an adversary can adaptively choose each
$f_i$ after seeing the
mechanism's first $i-1$ answers.  {\em This is the first interactive
mechanism better than the Laplace mechanism, and its performance is
close to the best possible even in the non-interactive setting.}

The basic implementation of the median mechanism is not efficient, and
we give an efficient implementation with a somewhat weaker utility
guarantee.  (The privacy guarantee is as strong as in the basic
implementation.)  This alternative implementation runs in time polynomial
in $n$, $k$, and $|X|$, and satisfies the following
(Theorem~\ref{thm:main2}): for every
sequence $f_1,\ldots,f_k$ of predicate queries, for all but a
negligible fraction of input distributions, the efficient median
mechanism is $(\eps,\delta)$-useful.

{\em This is the first efficient mechanism with a non-trivial utility
guarantee and polylogarithmic privacy cost, even in the
non-interactive setting.}

\subsection{The Main Ideas}

The key challenge to designing an interactive mechanism that
outperforms the Laplace mechanism lies in determining the appropriate
correlations between different output perturbations on the fly,
without knowledge of future queries.  It is not obvious that anything
significantly better than independent perturbations is possible in
the interactive setting.

Our median mechanism and our analysis of it can be summarized, at a
high level, by three facts.  First, among any set of $k$ queries,
we prove that there are $O(\log k \log |X|)$ ``hard'' queries,
the answers to which completely determine the answers to all of the
other queries (up to $\pm \eps$).  Roughly, this holds because:
(i) by a VC dimension argument, we can focus on databases over $X$ of
size only $O(\log k)$; and (ii) every time we answer a ``hard'' query,
the number of databases consistent with the mechanism's answers
shrinks by a constant factor, and this number cannot drop below~1
(because of the true input database).
Second, we design a method to privately release an indicator vector which distinguishes between hard and easy queries online. We note that a similar private `indicator vector' technique was used by Dwork et al. \cite{DNRRV09}.
Essentially, the median mechanism deems a query ``easy'' if a majority
of the databases that are consistent (up to $\pm \eps$) with the
previous answers of the mechanism would answer the
current query accurately.
The median mechanism answers the small number of hard queries using
independent Laplace perturbations.
It answers an easy query (accurately) using the
median query result given by databases that are consistent with
previous answers.
A key intuition is that if a user knows that query~$i$ is easy, {\em
then it can generate the mechanism's answer on its own}.  Thus
answering an easy query communicates only a single new bit of
information: that the query is easy.
Finally, we show how to release the classification of queries as
``easy'' and ``hard'' with low privacy cost; intuitively, this is
possible because (independent of the database) there can be only
$O(\log k \log |X|)$ hard queries.

Our basic implementation of the median mechanism is not efficient for
the same reasons as for the mechanism in~\cite{BLR08}: it requires
non-trivial sampling from a set of super-polynomial size.  For our
efficient implementation, we pass to {\em fractional} databases,
represented as fractional histograms with components indexed by $X$.
Here, we use the random walk technology of Dyer, Frieze, and
Kannan~\cite{DFK91} for convex bodies to perform efficient random
sampling.
To explain why our utility guarantee no longer holds for every input database, recall
the first fact used in the basic implementation: every answer to a hard
query shrinks the number of consistent databases by a constant factor,
and this number starts at $|X|^{O(\log k)}$ and cannot drop below~1.
With fractional databases (where polytope volumes play the
role of set sizes), the lower bound of~1 on the set of consistent
(fractional) databases no longer holds.
Nonetheless, we prove a lower bound on the volume of this set for
almost all fractional histograms (equivalently probability distributions), which salvages the $O(\log k \log |X|)$
bound on hard queries for databases drawn from such distributions.

\section{Preliminaries}\label{sec:prelim}

We briefly formalize the setting of the previous section and record
some important definitions.
We consider some finite domain $X$, and define a database $D$
to be an unordered set of elements from $X$ (with multiplicities
allowed).
We write $n = |D|$ to denote the size of the database.
We consider the set of Boolean
functions (predicates) $f:X\rightarrow \{0,1\}$.
We abuse notation and define a predicate query
$f(D):X^*\rightarrow
[0,1]$ as $|\{x \in D
: f(X) = 1\}|/|D|$, the function that computes the fraction of
elements of $D$ that satisfy predicate~$f$.
We say that an answer $a_i$ to a query $f_i$ is
{\em $\epsilon$-accurate} with respect to database $D$ if
$|f_i(D)-a_i| \leq \epsilon$.
A mechanism $M(D,(f_1,\ldots,f_k))$ is a function from
databases and
queries to distributions over outputs. In this paper, we consider
mechanisms that answer predicate queries numerically, and so the
range of our mechanisms is $\bR^k$.\footnote{From
$\eps$-accurate answers, one can efficiently reconstruct a synthetic
database that is consistent (up to $\pm \eps$) with those answers, if
desired~\cite{DNRRV09}.}

\begin{definition}\label{def:useful}
A mechanism $M$ is {\em $(\epsilon,\delta)$-useful} if
for every sequence of queries $(f_1,\ldots,f_k)$ and every
database $D$,
with probability at least $1-\delta$ it provides answers
$a_1,\ldots,a_k$ that are $\epsilon$-accurate for $f_1,\ldots,f_k$
and $D$.
\end{definition}

%Given a query, a mechanism induces a distribution over
%outputs. Intuitively, a mechanism satisfies the constraint of
Recall that differential privacy means that
changing the identity of a single element of the
input database does not affect the probability of any outcome by more
than a small factor.
Formally, given a database~$D$, we say that a database $D'$
of the same size is a \emph{neighbor} of $D$ if it differs in only a
single element: $|D \cap D'| = |D|-1$.

\begin{definition}\label{def:priv}
A mechanism $M$ satisfies {\em $(\alpha,\tau)$-differential privacy}
if for every subset $S \sse \bR^k$, every set of queries
$(f_1,\ldots,f_k)$, and every pair of neighboring databases
$D,D'$:
\begin{displaymath}
\Pr[M(D) \in S] \leq e^{\alpha} \cdot \Pr[M(D') \in S]  + \tau.
\end{displaymath}

%If $\delta = 0$, we say that a mechanism satisfies
%$\alpha$-differential privacy.
\end{definition}
\noindent
We are generally interested in the case where $\tau$ is a
{\em negligible function} of some of the problem parameters,
meaning one that goes to zero faster
than $x^{-c}$ for every constant $c$.

Finally, the {\em sensitivity} of a real-valued query is the largest
difference between its values on neighboring databases.  For example,
the sensitivity of every non-trivial predicate query is precisely~$1/n$.
%\begin{definition}[Negligible Function]\label{def:negligible}
%A function $f(m):\bR\rightarrow\bR$ is \emph{negligible} if $f(m) =
%o(\frac{1}{m^c})$ for every constant $c$.
%\end{definition}

%A non-interactive mechanism may be given the entire set of queries
%$(f_1,\ldots,f_k)$ up front. An interactive mechanism in contrast
%faces an online problem: an unknown set of queries arrives, and the
%answer to each query must be provided before the next query
%arrives.

%We study mechanisms which are $(\epsilon,\delta)$-useful and
%provide $(\alpha,\tau)$-differential privacy. $\epsilon$ and $\delta$
%will be fixed to hold independent of the number of queries
%$k$. Consequently, $\alpha$ will have to scale as some function of
%$k$. We provide an efficient interactive mechanism where this scaling
%factor is polylogarithmic in $k$, which is nearly optimal even for
%computationally unbounded non-interactive mechanisms.

\section{The Median Mechanism: Basic Implementation}\label{sec:alg}

We now describe the median mechanism and our basic implementation of
it.  As described in the Introduction, the mechanism is conceptually
simple.
It classifies queries as ``easy'' or ``hard'', essentially according
to whether or not a majority of the databases consistent with previous
answers to hard queries would give an accurate answer to it (in which
case the user already ``knows the answer'').
Easy queries are answered using the corresponding median value; hard
queries are answered as in the Laplace mechanism.

To explain the mechanism precisely,
we need to discuss a number of
parameters.  We take the privacy parameter $\alpha$, the accuracy
parameter $\eps$, and the number $k$ of queries as input; these are
hard constraints on the performance of our mechanism.\footnote{We
typically think of $\alpha,\eps$ as small constants, though our
results remain meaningful for some sub-constant values of $\alpha$
and $\eps$ as well.  We always assume that $\alpha$ is at least
inverse polynomial in~$k$.
Note that when $\alpha$ or $\eps$ is sufficiently small (at most $c/n$
for a small constant~$c$, say), simultaneously meaningful privacy and
utility is clearly impossible.}
Our mechanism obeys these constraints with a value of $\delta$
that is inverse polynomial in~$k$ and~$n$, and a value of $\tau$ that
is negligible in $k$ and $n$, provided $n$ is sufficiently large
(at least polylogarithmic in $k$ and $|X|$, see
Theorem~\ref{thm:main1}).
Of course, such a result can be rephrased as a nearly exponential
lower bound on the number of queries~$k$ that can be successfully
answered as a function of the database size~$n$.\footnote{In contrast,
the number of queries that the Laplace mechanism can privately and
usefully answer is at most linear.}

The median mechanism is shown in Figure~\ref{fig:mm}, and it makes use
of several additional parameters. For our analysis, we set their values to:
\begin{equation}\label{eq:m}
m = \frac{160000 \ln k\ln\frac{1}{\epsilon}}{\epsilon^2};
\end{equation}
\begin{equation}\label{eq:alpha'}
\alpha' = \frac{\alpha}{720m \ln |X|} =
\Theta\left(\frac{\alpha \epsilon^2}{\log |X|\log
k\log \frac{1}{\epsilon}}\right);
\end{equation}
\begin{equation}\label{eq:gamma}
\gamma = \frac{4}{\alpha' \eps n} \ln \frac{2k}{\alpha}
%\frac{20\ln k}{3\alpha'\epsilon}
=\Theta \left( \frac{\log |X| \log^2 k \log \tfrac{1}{\eps}}{\alpha
\eps^3 n}\right).
\end{equation}
The denominator in~\eqref{eq:alpha'} can be thought of as our
``privacy cost'' as a function of the number of queries~$k$.
Needless to say, we made no effort to optimize the constants.

The value~$r_i$ in Step~2(a) of the median mechanism is defined as
\begin{equation}\label{eq:r}
r_i = \frac{\sum_{S \in C_{i-1}
}\exp(-\eps^{-1}|f_i(D)-f_i(S)|)}{|C_{i-1}|}.
\end{equation}
For the Laplace perturbations in Steps~2(a) and~2(d), recall that the
distribution $\Lap(\sigma)$ has the cumulative distribution function
\begin{equation}\label{eq:laplace}
F(x) = 1-F(-x) = 1-\frac{1}{2} e^{-x/\sigma}.
\end{equation}

\begin{figure}[t]
\hrule\medskip
%\textbf{Input}:
%Algorithm: \alg
\begin{enumerate}

\item Initialize $C_0 = \{ \text{ databases of size $m$ over $X$ } \}$.

\item For each query $f_1,f_2,\ldots,f_k$ in turn:

\begin{enumerate}

\item Define $r_i$ as in~\eqref{eq:r} and
let $\hat{r}_i = r_i + \Lap(\tfrac{2}{\eps n \alpha'})$.

\item Let $t_i = \tfrac{3}{4} + j \cdot \gamma$, where $j \in \{
0,1,\ldots,\tfrac{1}{\gamma} \tfrac{3}{20} \}$ is chosen with probability
proportional to $2^{-j}$.

\item If $\hat{r}_i \ge t_i$, set $a_i$ to be the median value of
$f_i$ on $C_{i-1}$.

\item If $\hat{r}_i < t_i$, set $a_i$ to be $f_i(D) +
\Lap(\tfrac{1}{n \alpha'})$.

\item If $\hat{r}_i < t_i$, set $C_i$ to the databases $S$
of~$C_{i-1}$ with $|f_i(S)-a_i| \le \eps/50$; otherwise $C_i = C_{i-1}$.

\item If $\hat{r}_j < t_j$ for more than $20m \log |X|$ values of~$j
\le i$,
then halt and report failure.

\end{enumerate}

\end{enumerate}
\caption{\textsf{The Median Mechanism.}\label{fig:mm}}
\medskip\hrule
\end{figure}

The motivation behind the mechanism's steps is as follows.
The set $C_i$ is the set of size-$m$ databases consistent (up to $\pm
\eps/50$) with previous answers of the mechanism to hard queries.  The
focus on
databases with the small size $m$ is justified by a VC dimension
argument, see Proposition~\ref{uniformConvergence}.
Steps~2(a) and~2(b) choose a random value~$\hat{r}_i$ and a random
threshold~$t_i$.  The value $r_i$ in Step~2(a) is a measure of how
easy the query is, with higher numbers being easier.  A more obvious
measure would be the fraction of databases $S$ in $C_{i-1}$ for which
$|f_i(S)-f_i(D)| \le \eps$, but this is a highly sensitive statistic
(unlike $r_i$, see Lemma~\ref{sensitivityLemma}).  The mechanism uses
the perturbed value $\hat{r}_i$ rather than $r_i$ to
privately communicate which
queries are easy and which are hard.
In Step~2(b), we choose the threshold~$t_i$ at random between $3/4$
and $9/10$.  This randomly shifted threshold ensures that, for every
database~$D$, there is likely to be a significant gap between
$r_i$ and~$t_i$; such gaps are useful when optimizing the privacy
guarantee.
Steps~2(c) and~2(d) answer easy and hard queries, respectively.
Step~2(e) updates the set of databases consistent with previous
answers to hard queries.
We prove in Lemma~\ref{fewDiscounts} that Step~2(f) occurs with
at most inverse polynomial probability.

%Moreover, we will require that $n \geq \frac{(\log
%k)^2}{\alpha'\epsilon}$ --- a weak requirement that still allows the
%number of queries to be superpolynomial in the database size.

%XXX formula for Laplace noise

Finally, we note that the median mechanism is defined as if the total
number of queries $k$ is (approximately) known in advance.
This assumption can be removed by using successively doubling
``guesses'' of $k$; this increases the privacy cost by an
$O(\log k)$ factor.

\section{Analysis of Median Mechanism}\label{sec:analysis}

This section proves the following privacy and utility guarantees for
the basic implementation of the median mechanism.
\begin{theorem}\label{thm:main1}
For every sequence of adaptively chosen predicate queries
$f_1,\ldots,f_k$ arriving online, the median mechanism is
$(\epsilon,\delta)$-useful and $(\alpha,\tau)$-differentially private,
where $\tau$ is a negligible function of $k$ and $|X|$,
and $\delta$ is an inverse polynomial function of $k$ and $n$,
provided the database size~$n$ satisfies
\begin{equation}\label{eq:n}
n \ge \frac{30 \ln \tfrac{2k}{\alpha} \log_2 k}{\alpha'\epsilon} =
\Theta\left( \frac{\log |X|\log^3 k
\log \frac{1}{\epsilon}}{\alpha\epsilon^3}\right).
\end{equation}
\end{theorem}
We prove the utility and privacy guarantees in
Sections~\ref{subsec:util} and~\ref{subsec:priv},
respectively.\footnote{If desired, in Theorem~\ref{thm:main1}
we can treat $n$ as a parameter and solve for the error $\eps$.
The maximum error on any query (normalized by
the database size) is
$O(\log k \log^{1/3} |X|/n^{1/3}\alpha^{1/3})$; the unnormalized error
is a factor of~$n$ larger.}

%\subsection{Utility Guarantee for the Median
%Mechanism}\label{subsec:util}
\subsection{Utility of the Median
Mechanism}\label{subsec:util}

Here we prove a utility guarantee for the median mechanism.
%We think of the
%accuracy parameter $\eps$ as a fixed constant.
%We assume that the failure probability~$\delta$
%satisfies $\delta =
%k\exp(-\Omega(\epsilon n\alpha'))$, where $\alpha'$ is defined as
%in~\eqref{eq:alpha'}.
%, and is therefore negligible provided $\epsilon \alpha' = o(n)$.
%\newtheorem{theorem}{Theorem}
\begin{theorem}\label{thm:useful}
The median mechanism is $(\epsilon,\delta)$-useful, where
$\delta = k\exp(-\Omega(\epsilon n\alpha'))$.
\end{theorem}
Note that under assumption~\eqref{eq:n}, $\delta$ is inverse
polynomial in $k$ and $n$.

We give the proof of Theorem~\ref{thm:useful}
in three pieces: with high probability,
every hard query is answered accurately (Lemma~\ref{type2Accurate});
every easy query is answered accurately (Lemmas~\ref{accurater}
and~\ref{type1Accurate});
and the algorithm does not fail (Lemma~\ref{fewDiscounts}).
%The fact that there are only a logarithmic number of hard queries
%(Lemma~\ref{fewDiscounts}) also plays an important role in our privacy
%analysis (see Lemma~\ref{notManyBadLemma}).
%
The next two lemmas follow from the definition of the Laplace
distribution~\eqref{eq:laplace}, our choice of~$\delta$, and trivial
union bounds.
\begin{lemma}
\label{accurater}
With probability at least
$1-\tfrac{\delta}{2}$,
$|r_i - \hat{r}_i| \leq 1/100$ for every query $i$.
\end{lemma}

%\begin{proof}
%%The value $r_i$ is the average of $t$ independent random variables in
%%the range $[0,1]$. The Chernoff bound implies that $\Pr[|r_i -
%%\E[r_i]| \geq 1/200] \leq \exp(-\Omega(t)) = \delta/k$.
%The probability that $|r_i - \hat{r}_i| \geq
%1/100$ equals $\Pr[|\textrm{Lap}(\frac{2}{\epsilon n \alpha '})| \geq
%1/100]$, which from~\eqref{eq:laplace} is $\exp(-\Omega(\epsilon n
%\alpha'))$.
%%Combining these two facts with a
%A union bound over all $k$ queries proves the lemma.
%\end{proof}

\begin{lemma}
\label{type2Accurate}
With probability at least $1-\tfrac{\delta}{2}$,
every answer to a hard query is $(\epsilon/100)$-accurate for $D$.
\end{lemma}

%\begin{proof}
%The probability that the estimate of $f_i(D)$ is off by more than
%$\epsilon/100$ for a given hard query $i$ is
%$\Pr[\textrm{Lap}(\frac{1}{\alpha' n}) \geq \epsilon/100]$,
%which by~\eqref{eq:laplace} and our choice of $\delta$ is at most
%$\delta/2k$.  A union bound over the $k$ queries completes the proof.
%\end{proof}

The next lemma shows that median answers are accurate for easy queries.

\begin{lemma}
\label{type1Accurate}
If $|r_i - \hat{r}_i| \leq 1/100$ for every query $i$, then
every answer to an easy query is $\epsilon$-accurate for~$D$.
\end{lemma}

\begin{proof}
For a query $i$,
let $G_{i-1} = \{S \in C_{i-1} : |f_i(D)-f_i(S)| \leq
\epsilon\}$ denote the databases of $C_{i-1}$
on which the result of query~$f_i$ is
$\epsilon$-accurate for $D$.
Observe that if $|G_{i-1}| \ge .51 \cdot |C_{i-1}|$, then the median
value of~$f_i$ on~$C_{i-1}$ is an $\epsilon$-accurate answer
for~$D$.  Thus
proving the lemma reduces to showing that $\hat{r}_i \ge 3/4$ only
if $|G_{i-1}| \ge .51 \cdot |C_{i-1}|$.

Consider a query~$i$ with $|G_{i-1}| < .51 \cdot |C_{i-1}|$.
Using~\eqref{eq:r}, we have
%\begin{eqnarray*}
%\E[r_i]
\begin{eqnarray*}
r_i &=& \frac{\sum_{S \in
C_{i-1}}\exp(-\eps^{-1}|f_i(D)-f_i(S)|)}{|C_{i-1}|} \\
 &\leq& \frac{|G_{i-1}| + e^{-1}|C_{i-1} \sm G_{i-1}|}{|C_{i-1}|} \\
 &\leq& \frac{(\frac{51}{100}+\frac{49}{100e})|C_{i-1}|}{|C_{i-1}|} \\
 &<& \frac{74}{100}.
\end{eqnarray*}
%\end{eqnarray*}
%Lemma \ref{accurater} implies that the probability that $\hat{r}_i$ is
%at least $3/4$ for such a query is at most $\delta/2k$.  A union bound
%over the $k$ queries completes the proof.
Since $|r_i - \hat{r}_i| \leq 1/100$ for every query $i$ by
assumption, the proof is complete.
\end{proof}

Our final lemma shows that the median mechanism does not fail and
hence answers every query, with high probability; this will conclude
our proof of Theorem~\ref{thm:useful}.
We need the following preliminary proposition, which
instantiates the standard uniform convergence bound
with the fact that the VC dimension of every set of $k$
predicate queries is at most $\log_2 k$  \cite{Vap96}.
Recall the definition of the parameter~$m$ from~\eqref{eq:m}.
\begin{proposition}[Uniform Convergence Bound]
\label{uniformConvergence}
For every collection of $k$ predicate queries
$f_1,\ldots,f_k$ and every
database $D$, a database $S$ obtained by sampling points from $D$
uniformly at random will satisfy $|f_i(D)-f_i(S)| \leq \epsilon$ for
all $i$ except with probability $\delta$, provided
$$|S| \geq \frac{1}{2\epsilon^2} \left(\log k + \log
\frac{2}{\delta}\right).$$

In particular, there exists a database $S$ of size
$m$ such that for all $i \in \{1,\ldots,k\}$,
$|f_i(D)-f_i(S)| \leq \epsilon/400$.
\end{proposition}
In other words, the results of $k$ predicate queries on an arbitrarily
large database can be well approximated by those on a database
with size only $O(\log k)$.

\begin{lemma}
\label{fewDiscounts}
If $|r_i - \hat{r}_i| \leq 1/100$ for every query $i$ and
every answer to a hard query is $(\epsilon/100)$-accurate for $D$,
then the median mechanism answers fewer than $20 m\log |X|$ hard
queries (and hence answers all queries before terminating).
\end{lemma}

\begin{proof}
The plan is to track the contraction of $C_i$ as hard
queries are answered by the median mechanism.
Initially we have $|C_0| \le |X|^m$.
If the median mechanism answers a hard query~$i$,
then the definition of the mechanism and our hypotheses yield
$$
r_i \leq \hat{r}_i + \frac{1}{100} < t_i + \frac{1}{100} \leq \frac{91}{100}.
$$
We then claim that the size of the set $C_i = \{S \in C_{i-1} : |f_i(S)
- a_i| \leq \epsilon/50\}$ is at most $\tfrac{94}{100}|C_{i-1}|$.
For if not,
\begin{eqnarray*}
r_i & = & \frac{\sum_{S \in
C_{i-1}} \exp(-\eps^{-1}|f_i(S)-f_i(D)|)}{|C_{i-1}|}\\
& \ge & \frac{94}{100} \cdot \exp\left(-\frac{1}{50}\right) > \frac{92}{100},
\end{eqnarray*}
which is a contradiction.

Iterating now shows that the number of consistent
databases decreases exponentially with the number of hard queries:
\begin{equation}\label{eq:drop}
|C_k| \leq \left(\frac{94}{100}\right)^{h} |X|^m
\end{equation}
if $h$ of the $k$ queries are hard.

On the other hand, Proposition \ref{uniformConvergence} guarantees
the existence of
a database $S^* \in C_0$ for which $|f_i(S^*)-f_i(D)| \leq
\eps/100$ for every query~$f_i$.
Since all answers~$a_i$ produced by the median mechanism for hard
queries~$i$ are $(\eps/100)$-accurate for~$D$ by assumption,
$|f_i(S^*)-a_i| \leq |f_i(S^*)-f_i(D)| + |f_i(D)-a_i| \le \eps/50$.
This shows that $S^* \in C_k$ and hence $|C_k| \geq 1$.
Combining this with~\eqref{eq:drop} gives
$$
h  \leq \frac{m\ln |X|}{\ln (50/47)} < 20 m\ln |X|,
$$
as desired.
\end{proof}

%\begin{proof}[Proof of Theorem]
%The accuracy of all answered queries follows from lemmas
%\ref{type1Accurate} and \ref{type2Accurate}. By lemma
%\ref{fewDiscounts}, the algorithm does not abort and so answers all
%queries.
%\end{proof}

%\subsection{Privacy Guarantee for the Median Mechanism}\label{subsec:priv}
\subsection{Privacy of the Median Mechanism}\label{subsec:priv}

This section establishes the following privacy guarantee for the
median mechanism.
\begin{theorem}\label{thm:priv}
The median mechanism is $(\alpha,\tau)$- differentially private,
%$\alpha = O(\frac{\log |X|\log
%k\log(1/\epsilon)}{\epsilon^2}\alpha')$ and
where $\tau$ is a negligible function of $|X|$ and $k$
when $n$ is sufficiently large (as in~\eqref{eq:n}).
\end{theorem}

We can treat the median mechanism as if it has two outputs: a vector
of answers $\vec{a} \in \bR^k$, and a vector $\vec{d} \in \{0,1\}^k$
such that $d_i = 0$ if $i$ is an easy query and $d_i = 1$ if $i$ is a
hard query.
A key observation in the privacy analysis is that
answers to easy queries are a function only of the previous output of the
mechanism, and incur no additional privacy cost beyond the release of
the bit $d_i$.
Moreover, the median mechanism is guaranteed to produce
no more than $O(m\log |X|)$ answers to hard queries.
Intuitively, what we need to show is that the vector $\vec{d}$ can be
released after an unusually small perturbation.
%We write $\vec{d}_i$ and $\vec{a}_i$ to refer to the
%prefixes of length $i$ of vectors $\vec{d}$ and $\vec{a}$
%respectively.

%We prove a result stronger than Theorem~\ref{thm:priv}: for
%every input database~$D$ and queries $f_1,\ldots,f_k$ chosen by an
%adaptive adversary, with probability at least $1-\tau$ over the
%internal randomness of the median mechanism, the mechanism is
%$(\alpha,0)$-differentially private.

%The failure probability $\tau$ corresponds to two highly
%unlikely events.
%%First, let $\e_1$ denote the event that the median mechanism
%%answers at most $20m \ln |X|$ hard queries.  Lemma~\ref{fewDiscounts}
%%proves that $\Pr[\e_1]$ is negligible.
%%Second
%First, let $\e_1$ denote the
%event that one of the Laplace random variables sampled
%in Step~2(a) of the median mechanism takes on a value at least
%$\tfrac{\gamma}{2} - \tfrac{2}{\eps n}$.
%The definition of the Laplace distribution~\eqref{eq:laplace} and our
%choice of~$\gamma$~\eqref{eq:gamma} immediately imply that $\Pr[\e_1]$
%is negligible.

%Second, for a given database $D$ and queries $f_1,\ldots,f_k$ chosen
%by an adaptive adversary, we say that the threshold $t_i$
%chosen by the median mechanism in Step~2(b) is
%\emph{high} if $r_i < t_i - \tfrac{\gamma}{2}$, \emph{borderline} if $r_i \in
%[t_i - \tfrac{\gamma}{2}, t_i + \tfrac{\gamma}{2}]$, and {\em low} if
%$r_i > t_i + \tfrac{\gamma}{2}$.
%Let $\e_2$ denote the event that more than $80 m \ln n$ theresholds
%are borderline.

Our first lemma states that the small sensitivity of predicate queries
carries over, with a $2/\eps$ factor loss, to the $r$-function defined
in~\eqref{eq:r}.

\begin{lemma}
\label{sensitivityLemma}
The function $r_i(D) = (\sum_{S \in C}\exp(-\eps^{-1}|f(D)-f(S)|)/|C|$
has sensitivity $\frac{2}{\epsilon n}$
for every fixed set $C$ of databases and predicate query~$f$.
\end{lemma}

\begin{proof}
Let $D$ and $D'$ be neighboring databases. Then
\begin{eqnarray*}
r_i(D)
& = &
\frac{\sum_{S \in C}\exp(-\eps^{-1}|f(D)-f(S)|)}{|C_i|}\\
& \leq &\frac{\sum_{S\in
C_i}\exp(-\eps^{-1}(|f(D')-f(S)|-n^{-1}))}{|C_i|} \\
& = &  \exp \left(\frac{1}{\epsilon n} \right)\cdot r_i(D') \\
& \leq & \left(1+\frac{2}{\epsilon n}\right) \cdot r_i(D') \\
& \leq & r_i(D') + \frac{2}{\epsilon n}
\end{eqnarray*}
where the first inequality follows from the fact that the (predicate)
query $f$ has sensitivity $1/n$,
the second from the fact that $e^{x} \le 1+2x$ when $x \in [0,1]$,
and the third from the fact that $r_i(D') \leq 1$.
\end{proof}

The next lemma identifies nice properties of ``typical executions'' of
the median mechanism.
Consider an output $\out$ of the median mechanism with a database~$D$.
From $D$ and $\out$, we can uniquely recover the values
$r_1,\ldots,r_k$ computed (via~\eqref{eq:r}) in Step~2(a) of the
median mechanism, with $r_i$ depending only on the first $i-1$
components of $d$ and $a$.
We sometimes write such a value as $r_i(D,\out)$, or as $r_i(D)$
if an output $\out$ has been fixed.
Call a possible threshold $t_i$ {\em good} for $D$ and $\out$ if
$d_i = 0$ and $r_i(D,\out) \ge t_i + \gamma$, where $\gamma$ is
defined as in~\eqref{eq:gamma}.
Call a vector $\vec{t}$ of possible thresholds {\em good} for $D$
and $\out$ if all but $180m \ln |X|$ of the thresholds are good for
$D$ and $\out$.

\begin{lemma}
\label{notManyBadLemma}
For every database $D$, with all but
negligible ($\exp(-\Omega(\log k\log |X|/\epsilon^2))$) probability,
the thresholds $\vec{t}$ generated by the median mechanism are good for its
output $\out$.
\end{lemma}

\begin{proof}
The idea is to ``charge'' the probability of bad thresholds to
that of answering hard queries, which are strictly limited by the
median mechanism.
%we already bounded in Lemma~\ref{fewDiscounts}.
Since the median mechanism only allows $20m \ln |X|$ of the $d_i$'s to be~1, we
only need to bound the number of queries~$i$ with output $d_i = 0$ and
threshold $t_i$ satisfying $r_i < t_i + \gamma$, where $r_i$ is the
value computed by the median mechanism in Step~2(a) when it answers
the query~$i$.

Let~$Y_i$ be the indicator random variable corresponding to the
(larger) event that $r_i < t_i + \gamma$.
Define $Z_i$ to be~1 if and only if, when answering the $i$th query,
the median mechanism chooses a threshold $t_i$ and a Laplace perturbation
$\Delta_i$ such that $r_i + \Delta_i < t_i$ (i.e., the query is
classified as hard).
%We make one exception to the definitions: i
If the median mechanism
fails before reaching query~$i$, then we define $Y_i = Z_i = 0$.
Set $Y = \sum_{i=1}^k Y_i$ and $Z = \sum_{i=1}^k Z_i$.
We can finish the proof by showing that $Y$ is at most $160m \ln |X|$
except with negligible probability.

Consider a query~$i$ and condition on the event that
%the past, which fixes~$r_i$.
%First suppose that
$r_i \ge \tfrac{9}{10}$; this event depends only on the results of
previous queries.
In this case, $Y_i = 1$ only if $t_i = 9/10$.
But this occurs with probability $2^{-3/20\gamma}$, which
using~\eqref{eq:gamma} and~\eqref{eq:n} is at most~$1/k$.\footnote{For
simplicity, we ignore the normalizing constant in the distribution
over $j$'s in Step~2(b), which is $\Theta(1)$.}
%$$
%2^{-3/20\gamma} = 2^{-\alpha'\epsilon n/\log k} \leq \frac{1}{k},
%$$
%where the equality follows from~\eqref{eq:gamma} and
%the inequality from~\eqref{eq:n}.
Therefore, the expected contribution to~$Y$
coming from queries~$i$ with $r_i \ge \tfrac{9}{10}$ is at most $1$.
Since~$t_i$ is selected independently at random for each $i$, the
Chernoff bound implies that the probability that such queries
contribute more than $m\ln |X|$ to $Y$ is
$$\exp(-\Omega((m \log |X|)^2)) =
\exp(-\Omega((\log k)^2(\log |X|)^2/\epsilon^4)).$$

%Next consider a query~$i$ with $r_i < \tfrac{9}{10}$.
Now condition on the event that $r_i < \tfrac{9}{10}$.
Let $T_i$ denote the threshold choices that would cause $Y_i$ to be~1,
and let $s_i$ be the smallest such;
since $r_i < \tfrac{9}{10}$, $|T_i| \ge 2$.
For every $t_i \in T_i$, $t_i > r_i - \gamma$; hence, for every
$t_i \in T_i \sm \{s_i\}$, $t_i > r_i$.
Also, our distribution on the $j$'s in Step~2(b) ensures that
$\Pr[t_i \in T_i \sm \{s_i\} ] \ge \tfrac{1}{2} \Pr[t_i \in T_i]$.
%We claim that $\E[Y_i] \le 4\E[Z_i]$.
%, conditioned arbitrarily on the
%past and hence also unconditionally (subject to $r_i < \tfrac{9}{10}$).
%If the median mechanism fails before query~$i$ this is clearly true,
%so suppose this is not the case.
Since the Laplace distribution is symmetric around zero and the
random choices $\Delta_i,t_i$ are independent, we have
\begin{eqnarray}\nonumber
\E[Z_i] & = & \Pr[t_i > r_i + \Delta_i]\\ \nonumber
& \ge & \Pr[t_i > r_i] \cdot \Pr[\Delta_i \le 0]\\ \nonumber
& \ge & \tfrac{1}{4} \Pr[t_i > r_i - \gamma]\\ \label{eq:charge}
& = & \tfrac{1}{4} \E[Y_i].
\end{eqnarray}
The definition of the median mechanism ensures that $Z \le 20m \ln |X|$ with
probability~1.
%Now, Lemma~\ref{fewDiscounts} implies that $\E[Z] \le 20m \ln
%|X|$.\footnote{That lemma proves a high probability result;
%the argument can be modified to give the expectation bound used here.}
Linearity of expectation, inequality~\eqref{eq:charge}, and the
 Chernoff bound
% imply
%that $\E[Y] \le 80 m \ln |X|$.  The Chernoff bound implies that
%$Y \le
imply that queries with~$r_i < \tfrac{9}{10}$ contribute at most
$159 m \ln |X|$ to~$Y$ with probability at least $1-\exp(-\Omega(\log k\log
|X|/\epsilon^2))$.
%, completing the proof.
The proof is complete.
\end{proof}

We can now prove Theorem~\ref{thm:priv}.

%\begin{lemma}
%The vectors $\vec{d}$ and $\vec{a}$ are released preserving
%$(\alpha,\tau)$-differential privacy, where $\alpha = O(\frac{\log
%|X|\log k\log(1/\epsilon)}{\epsilon^2}\alpha')$ and $\tau =
%\tau(k,|X|)$ is a negligible function of $k$ and $|X|$.
%\end{lemma}

\vspace{.1in}

\begin{prevproof}{Theorem}{thm:priv}
Recall Definition~\ref{def:priv} and fix a database~$D$,
queries $f_1,\ldots,f_k$, and a subset $S$ of possible mechanism
outputs.  For simplicity, we assume that all perturbations are drawn
from a discretized Laplace distribution, so that the median mechanism
has a countable
range; the continuous case can be treated using similar arguments.
Then, we can think of~$S$ as a countable set of output vector pairs
$\out$ with $d \in \{0,1\}^k$ and $a \in \bR^k$.
%as partitioned according to
%which queries are deemed easy or hard by the
%median mechanism (summarized as
%an effective output vector $\vec{d} \in \{0,1\}^k$).  For the class
%of the partition corresponding to~$d$, there are corresponding ranges
%$R(d) = (R_1(\vec{d}),\ldots,R_k(\vec{d}))$ of (real-valued) answers to the
%$k$ queries.
We write $MM(D,\vec{f}) = \out$
%(\vec{d},\vec{R}(\vec{d}))$
for the event that the median mechanism classifies
the queries $f=(f_1,\ldots,f_k)$ according
to~$\vec{d}$ and outputs the numerical answers $\vec{a}$.
% in the corresponding ranges
%$\vec{R}(\vec{d})$.
If the mechanism computes thresholds $\vec{t}$
while doing so, we write
$MM(D,\vec{f}) = (t,d,a)$.
%(\vec{t},\vec{d},\vec{R}(\vec{d}))$.
Let $G(\out,D)$ denote the vectors that would be good thresholds for
$\out$ and~$D$.  (Recall that $D$ and $\out$ uniquely
define the corresponding $r_i(D,\out)$'s.)

We have
$$
\Pr[ MM(D,\vec{f}) \in S] = \sum_{\out \in S} \Pr[MM(D,\vec{f}) = \out]
$$
$$
\le \tau + \sum_{\out \in S} \Pr[MM(D,\vec{f})= (\vec{t},\vec{d},a)]
$$
$$
= \tau + \sum_{\out \in S} \sum_{t \in G(\out,D)}
\Pr[MM(D,\vec{f}) = (\vec{t},\vec{d},a)]
$$
with some $\vec{t}$ good for $\out,D$, and where $\tau$ is the
negligible function of Lemma~\ref{notManyBadLemma}.
We complete the proof by showing that,  for every neighboring
database~$D'$, possible output $\out$, and
thresholds $t$ good for $\out$ and $D$,
\begin{equation}\label{eq:main}
\Pr[MM(D,\vec{f}) = (\vec{t},\vec{d},a)]
\le
e^{\alpha} \cdot \Pr[MM(D',\vec{f}) = (\vec{t},\vec{d},a)].
\end{equation}

Fix a neighboring database~$D'$, a target output $\out$,
and thresholds~$t$ good for $\out$ and $D$.
The probability that the median mechanism chooses the target thresholds~$t$ is
independent of the underlying database, and so is the same on both
sides of~\eqref{eq:main}.  For the rest of the proof, we condition on
the event that the median mechanism uses the thresholds $t$ (both with
database $D$ and database $D'$).

Let $\e_i$ denote the event that~$MM(D,f)$ classifies the first $i$
queries in agreement with the target output (i.e., query $j \le i$
is deemed easy if and only if $d_j=0$)
and that its first $i$ answers are $a_1,\ldots,a_i$.
Let $\e'_i$ denote the analogous event for $MM(D',f)$.
Observe that $\e_k,\e'_k$ are the relevant events on the left- and
right-hand sides of~\eqref{eq:main}, respectively (after conditioning
on~$t$).
If $\out$ is such that the median mechanism would fail after the
$\ell$th query, then
the following proof should be applied to $\e_{\ell},\e'_{\ell}$
instead of $\e_k,\e'_k$.
We next give a crude upper bound on the ratio $\Pr[\e_i |
\e_{i-1}]/\Pr[\e'_i | \e'_{i-1} ]$ that holds for every query
(see~\eqref{eq:crude}, below), followed by a much better upper bound
for queries with good thresholds.

Imagine running the median mechanism in parallel on $D,D'$ and
condition on the events $\e_{i-1},\e'_{i-1}$.
The set $C_{i-1}$ is then the same in both runs of the mechanism,
and $r_i(D),r_i(D')$ are now fixed.
Let $b_i$ ($b'_i$) be 0 if $MM(D,f)$ ($MM(D',f)$) classifies query~$i$
as easy and~1 otherwise.
Since $r_i(D') \in [r_i(D) \pm \tfrac{2}{\eps n}]$
(Lemma~\ref{sensitivityLemma}) and
a perturbation with distribution $\Lap(\tfrac{2}{\alpha' \eps n})$
is added to these values before comparing to the threshold~$t_i$
(Step~2(a)),
$$\Pr[ b_i = 0 \,|\, \e_{i-1} ]
\le e^{\alpha'} \Pr[ b'_i = 0 \,|\, \e'_{i-1} ]$$
and similarly
for the events where $b_i,b'_i = 1$.
Suppose that the target classification is $d_i=1$ (a hard query),
and let $s_i$ and $s'_i$ denote the random variables
$f_i(D) + \Lap(\tfrac{1}{\alpha'n})$ and $f_i(D') +
\Lap(\tfrac{1}{\alpha'n})$, respectively.
Independence of the Laplace perturbations
in Steps~2(a) and~2(d) implies that
$$\Pr[\e_i | \e_{i-1}] = \Pr[ b_i = 1 \,|\, \e_{i-1}] \cdot
\Pr[ s_i = a_i \,|\, \e_{i-1}]$$
and
$$\Pr[\e'_i | \e'_{i-1}] = \Pr[ b'_i = 1 \,|\, \e'_{i-1}] \cdot
\Pr[ s'_i = a_i \,|\, \e'_{i-1}].$$
Since the predicate query $f_i$
has sensitivity $1/n$, we have
\begin{equation}\label{eq:crude}
\Pr[\e_i \,|\, \e_{i-1}] \le e^{2\alpha'} \cdot \Pr[\e'_i \,|\, \e'_{i-1}]
\end{equation}
when $d_i=1$.

Now suppose that $d_i=0$, and
let $m_i$ denote the median value of $f_i$ on $C_{i-1}$.
Then $\Pr[\e_i | \e_{i-1}]$ is either~0 (if $m_i \neq a_i$) or
$\Pr[ b_i = 0 \,|\, \e_{i-1} ]$ (if $m_i = a_i$);
similarly, $\Pr[\e'_i | \e'_{i-1}]$ is either~0 or
$\Pr[ b'_i = 0 \,|\, \e'_{i-1} ]$.
Thus the bound in~\eqref{eq:crude} continues to hold (even with
$e^{2\alpha'}$ replaced by $e^{\alpha'}$) when $d_i=0$.

Since $\alpha'$ is not much smaller than the privacy target~$\alpha$
(recall~\eqref{eq:alpha'}), we cannot afford to suffer the upper bound
in~\eqref{eq:crude} for many queries.  Fortunately, for queries~$i$
with good thresholds we can do much better.
Consider a query~$i$ such that $t_i$ is good for $\out$ and $D$ and
condition again on $\e_{i-1},\e'_{i-1}$, which fixes $C_{i-1}$ and
hence $r_i(D)$.
Goodness implies that $d_i=0$, so the arguments from the previous
paragraph also apply here.  We can therefore assume that the median
value $m_i$ of $f_i$ on $C_{i-1}$ equals $a_i$ and focus on bounding
$\Pr[ b_i = 0 \,|\, \e_{i-1} ]$ in terms of $\Pr[ b'_i = 0 \,|\, \e'_{i-1} ]$.
Goodness also implies that $r_i(D) \ge t_i + \gamma$ and hence
$r_i(D') \ge t_i + \gamma - \tfrac{2}{\eps n} \ge t_i +
\tfrac{\gamma}{2}$ (by Lemma~\ref{sensitivityLemma}).
Recalling from~\eqref{eq:gamma} the definition of~$\gamma$, we have
\begin{eqnarray}\nonumber
\Pr[ b'_i = 0 \,|\, \e'_{i-1} ]
& \ge & \Pr[r_i - \hat{r}_i < \tfrac{\gamma}{2}]\\ \nonumber
& = & 1 - \tfrac{1}{2}e^{-\gamma \alpha' \eps n/4}\\ \label{eq:sharp}
& = & 1 - \frac{\alpha}{4k}
\end{eqnarray}
and of course, $\Pr[ b_i = 0 \,|\, \e_{i-1} ] \le 1$.

Applying~\eqref{eq:crude} to the bad queries --- at most $180m \ln
|X|$ of them, since $t$ is good for $\out$ and $D$ ---
and~\eqref{eq:sharp} to the rest, we can derive
\begin{eqnarray*}
\Pr[\e_k] &=& \prod_{i=1}^k \Pr[\e_i \,:\, \e_{i-1}] \\
& \le &
\underbrace{e^{360\alpha' m \ln |X|}}_{\le e^{\alpha/2} \text{ by~\eqref{eq:alpha'}}} \cdot
\underbrace{(1 - \frac{\alpha}{4k})^{-k}}_{\le
(1+\tfrac{\alpha}{2k})^k \le e^{\alpha/2}} \cdot
\prod_{i=1}^k \Pr[\e'_i \,:\, \e'_{i-1}] \\
&\le& e^{\alpha} \cdot \Pr[\e'_k],
\end{eqnarray*}
which completes the proof of both the inequality~\eqref{eq:main} and
the theorem.
\end{prevproof}

\section{The Median Mechanism:\\ Efficient Implementation}
\label{efficientSection}

The basic implementation of the median mechanism runs in time
$|X|^{\Theta(\log k\log(1/\epsilon)/\epsilon^2)}$.  This section
provides an efficient implementation, running in time polynomial in
$n$, $k$, and $|X|$, although with a weaker usefulness guarantee.
\begin{theorem}\label{thm:main2}
Assume that the database size~$n$ satisfies~\eqref{eq:n}.
For every sequence of adaptively chosen predicate queries
$f_1,\ldots,f_k$ arriving online, the
efficient implementation of the median Mechanism is
$(\alpha,\tau)$-differentially private for a negligible
function~$\tau$.
Moreover, for every fixed set $f_1,\ldots,f_k$ of queries,
it is $(\epsilon,\delta)$-useful for all but a negligible fraction of \emph{fractional} databases (equivalently, probability distributions).
\end{theorem}
Specifically, our mechanism answers exponentially many queries for all
but an $O(|X|^{-m})$ fraction of probability distributions over $X$ drawn from the unit $\ell_1$ ball, and from databases drawn from such distributions.
Thus our efficient implementation always guarantees privacy,
but
%promises usefulness only for almost all databases $D$.
for
%any particular
a given set of queries $f_1,\ldots,f_k$,
there might be a negligibly small fraction of fractional histograms for which our
mechanism is not useful for all $k$ queries.

We note however that even for the small fraction of fractional histograms for which the efficient median mechanism may not satisfy our usefulness guarantee, it does not output incorrect answers: it merely halts after having answered a sufficiently large number of queries using the Laplace mechanism. Therefore, even for this small fraction of databases, the efficient median mechanism is an improvement over the Laplace mechanism: in the worst case, it simply answers every query using the Laplace mechanism before halting, and in the best case, it is able to answer many more queries.

%Notice
%that this negligible fraction of databases is defined with respect to
%a particular set of queries, so the queries to the efficient
%Consistency Mechanism should be regarded as being chosen by a
%\emph{non-adaptive} adversary.

%Consider the run time of the : except for sampling
%elements uniformly at random from $C_{i-1}$ and constructing set $C_i$
%from set $C_{i-1}$, for each query $i$, the mechanism simply performs
%a polynomial number of elementary operations. Unfortunately, $|C_0| =
%\Omega(|X|^m)$, which is superpolynomial in $|X|$ and $k$. Sampling
%from any $C_i$ naively takes time polynomial in $|C_i|$, and therefore
%the consistency mechanism as presented does not run in time
%poly$(|X|,k)$. In this section, we

We give a high-level overview of the proof
of Theorem~\ref{thm:main2} which we then make formal.
First, why isn't the median mechanism a computationally efficient mechanism?
Because~$C_0$ has super-polynomial size~$|X|^m$, and computing~$r_i$
in Step~2(a), the median value in Step~2(c), and the set~$C_i$ in
Step~2(e) could require time proportional to $|C_0|$.  An obvious idea
is to randomly sample elements of $C_{i-1}$ to
approximately compute $r_i$
and the median value of $f_i$ on $C_{i-1}$; while it is easy to
control the resulting sampling error and preserve the utility and
privacy guarantees of Section~\ref{sec:analysis},  it is not clear how
to sample from $C_{i-1}$ efficiently.

%This section
We show how to implement the
median mechanism in polynomial time by redefining the sets $C_i$ to
be sets of probability distributions over points in $X$ that are
consistent (up to $\pm \frac{\eps}{50}$) with the hard queries answered
up to  the $i$th query. Each set $C_i$ will be a convex polytope in
$\bR^{|X|}$ defined by the intersection of at most $O(m\log |X|)$
halfspaces, and hence it will be possible to sample points from $C_i$
approximately uniformly at random in time poly$(|X|,m)$ via the
grid walk of Dyer, Frieze, and Kannan \cite{DFK91}.
Lemmas \ref{accurater}, \ref{type2Accurate}, and
\ref{type1Accurate} still hold (trivially modified to accommodate
sampling error).  We have to
reprove Lemma~\ref{fewDiscounts}, in a somewhat weaker form:
that for all but a diminishing fraction of input databases $D$, the
median mechanism does not abort except with probability
$k\exp(-\Omega(\epsilon n\alpha'))$.
%We are unfortunately not
%able to make such a claim for \emph{all} input databases $D$, as we
%are for the inefficient implementation.
As for our privacy analysis of the median mechanism, it is independent
of the representation of the sets $C_i$ and the mechanisms' failure
probability, and so it need not be repeated --- the efficient
implementation is provably private for {\em all} input databases and
query sequences.

We now give a formal analysis of the efficient implementation.
\subsection{Redefining the sets $C_i$}

We redefine the sets $C_i$ to represent databases that can
contain points fractionally, as opposed to the finite set of
small discrete databases.
Equivalently, we can view the sets $C_i$ as containing probability
distributions over the set of points $X$.

We initialize $C_0$ to be the $\ell_1$ ball of radius $m$ in
$\bR^{|X|}$, $m B_1^{|X|}$, intersected with the non-negative orthant:
$$
C_0 = \{\vec{F} \in \bR^{|X|} : \vec{F} \geq 0, ||\vec{F}||_1 \leq m\}.
$$
Each dimension $i$ in $\bR^{|X|}$ corresponds
to an element $x_i \in X$. Elements $\vec{F} \in C_0$ can be viewed
as fractional histograms.
% corresponding to databases $D \subset X$.
Note that integral points in $C_0$ correspond exactly to databases
of size at most $m$.

We generalize our query functions $f_i$ to fractional histograms in the
natural way:
%$ to be defined over points
%$\vec{F} \in C_0$ in the natural way:
$$
f_i(\vec{F}) = \frac{1}{m}\sum_{j : f_i(x_j) = 1}\vec{F}_j.
$$

The update operation after a hard query $i$ is answered is the same
as in the basic implementation:
$$
C_{i} \leftarrow \left\{\vec{F} \in C_{i-1} : |f_i(\vec{F}) - a_i| \leq
\frac{\eps}{50}\right\}.
$$
Note that each updating operation after a hard query merely
intersects $C_{i-1}$ with the pair of halfspaces:
$$
\sum_{j : f_i(x_j) = 1}\vec{F}_j \leq m a_i + \frac{\eps m}{50}
\qquad \text{and} \qquad
\sum_{j : f_i(x_j) = 1}\vec{F}_j \geq m a_i - \frac{\eps m}{50};
$$
and so $C_{i}$ is a convex polytope for each $i$.

Dyer, Kannan, and Frieze~\cite{DFK91} show how to $\delta$-approximate
a random sample from a convex body $K \in \bR^{|X|}$ in time polynomial in
$|X|$ and the running time of a membership oracle for $K$, where
$\delta$ can be taken to be exponentially small (which is more than
sufficient for our purposes). Their algorithm has two requirements:
\begin{enumerate}

\item There must be an efficient membership oracle which can in
polynomial time determine whether a point $\vec{F}\in \bR^{|X|}$ lies
in $K$.

\item $K$ must be `well rounded':  $B_2^{|X|} \subseteq K \subseteq
|X|B_2^{|X|}$, where $B_2^{|X|}$ is the unit $\ell_2$ ball in
$\bR^{|X|}$.

\end{enumerate}
Since $C_i$ is given as the intersection of a set of explicit
halfspaces, we have a simple membership oracle to determine whether a
given point $\vec{F} \in C_i$: we simply check that $\vec{F}$ lies
on the appropriate side of each of the halfspaces. This takes time
poly$(|X|,m)$, since the number of halfspaces defining $C_i$ is linear
in the number of answers to hard queries
given before time $i$, which is never
more than $20m \ln |X|$. Moreover, for each $i$ we have
$
C_i \subseteq C_0 \subset m B_1^{|X|} \subset  m B_2^{|X|} \subset
|X| B_2^{|X|}.
$
Finally, we can safely assume that $B_2^{X} \subseteq
C_i$ by simply considering the convex set $C'_i = C_i + B_2^{X}$
instead. This will not affect our results.

Therefore, we can implement the median mechanism in time
poly$(|X|,k)$ by using sets $C_i$ as defined in this section, and
sampling from them using the grid walk of \cite{DFK91}.
Estimation error in computing $r_i$ and the median value of $f_i$
on $C_{i-1}$ by random sampling rather than brute force is easily
controlled via the Chernoff bound and can be incorporated into
the proofs of Lemmas~\ref{accurater} and~\ref{type1Accurate} in the
obvious way.  It remains to
prove a continuous version of Lemma \ref{fewDiscounts}
to show that the efficient implementation of the median mechanism
is $(\epsilon,\delta)$-useful on all but a negligibly small fraction
of fractional histograms $F$.

\subsection{Usefulness for Almost All Distributions}

We now prove an analogue of Lemma~\ref{fewDiscounts} to
establish a usefulness guarantee for the efficient version of the
median mechanism.

\begin{definition}
With respect to any set of $k$ queries $f_1,\ldots,f_k$ and for any
$\vec{F}^* \in C_0$, define
$$
\textrm{Good}_{\epsilon}(\vec{F}^*) =
\{\vec{F} \in C_0 : \max_{i\in \{1,2,\ldots,k\}}
|f_i(\vec{F})-f_i(\vec{F}^*)| \leq \epsilon\}
$$
as the set of points that agree up to an additive $\epsilon$ factor
with $\vec{F}^*$ on every query $f_i$.

Since databases $D \subset X$ can be identified with their
corresponding histogram vectors $\vec{F} \in \bR^{|X|}$, we can also
write $\textrm{Good}_\epsilon(D)$ when the meaning is clear from
context.
\end{definition}
For any $\vec{F}^*$, $\textrm{Good}_\epsilon(\vec{F}^*)$ is a convex
polytope contained inside $C_0$.
We will prove that the efficient version of the median mechanism
is $(\eps,\delta)$-useful for a database $D$ if
\begin{equation}\label{eq:volume}
\frac{\Vol(\textrm{Good}_{\epsilon/100}(D))}{\Vol(C_0)} \geq
\frac{1}{|X|^{2m}}.
\end{equation}
We first prove that~\eqref{eq:volume} holds for almost every
fractional histogram.  For this, we need a preliminary lemma.

\begin{lemma}
\label{spacePartitionLemma}
Let $\mathcal{L}$ denote the set of integer points inside $C_0$. Then
with respect to an arbitrary set of $k$ queries,
$$
C_0 \subseteq \bigcup_{\vec{F}\in \mathcal{L}}
\textrm{Good}_{\epsilon/400}(\vec{F}).
$$
\end{lemma}

\begin{proof}
%This follows immediately from the uniform convergence bound
%(Proposition~\ref{uniformConvergence}).
Every rational valued point $\vec{F} \in
C_0$ corresponds to some (large) database $D \subset X$ by scaling
$\vec{F}$ to an integer-valued histogram. Irrational points can be
arbitrarily approximated by such a finite database.
By Proposition~\ref{uniformConvergence}, for every set of $k$
predicates $f_1,\ldots,f_k$, there is a database $F^* \subset X$
with $|F^*| = m$ such that for each $i$, $|f_i(F^*)-f_i(F)| \leq
\epsilon/400$.
Recalling that the histograms corresponding to databases of size at
most $m$ are exactly the integer points in $C_0$, the proof is complete.
\end{proof}

\begin{lemma}
\label{goodVolumeLemma}
All but an $|X|^{-m}$ fraction of \emph{fractional} histograms $F$
satisfy
$$\frac{\Vol(\textrm{Good}_{\epsilon/200}(F))}{\Vol(C_0)} \geq
\frac{1}{|X|^{2m}}.$$
%are such that:
%$$
%\frac{\Vol(\textrm{Good}_{\epsilon/100}(F))}{\Vol(C_0)} \geq
%\frac{1}{|X|^{2m}}
%$$
\end{lemma}

\begin{proof}
Let
$$
\mathcal{B} = \left\{ F \in \mathcal{L} :
\frac{\Vol(\textrm{Good}_{\epsilon/400}(F))}{\Vol(C_0)} \leq
\frac{1}{|X|^{2m}} \right\}.
$$
Consider a randomly selected fractional histogram $F^* \in C_0$.
For any $F \in \mathcal{B}$ we have:
$$
\Pr[F^* \in \textrm{Good}_{\epsilon/400}(F)] =
\frac{\Vol(\textrm{Good}_{\epsilon/400}(F))}{\Vol(C_0)} <
\frac{1}{|X|^{2m}}
$$
Since $|\mathcal{B}| \leq |\mathcal{L}| \leq |X|^m$, by a union bound
we can conclude that except with probability $\frac{1}{|X|^{m}}$, $F^*
\not\in \textrm{Good}_{\epsilon/400}(F)$ for any $F \in
\mathcal{B}$.
However, by Lemma~\ref{spacePartitionLemma}, $F^* \in
\textrm{Good}_{\epsilon/400}(\vec{F'})$ for some $F' \in
\mathcal{L}$.
Therefore, except with probability $1/|X|^m$,
$F' \in \mathcal{L}\setminus \mathcal{B}$.
Thus, since
$\textrm{Good}_{\epsilon/400}(\vec{F'}) \subseteq
\textrm{Good}_{\epsilon/200}(\vec{F^*})$,
except with negligible probability, we have:
$$
\frac{\Vol(\textrm{Good}_{\epsilon/200}(F^*))}{\Vol(C_0)} \geq
\frac{\Vol(\textrm{Good}_{\epsilon/400}(F'))}{\Vol(C_0)} \geq
\frac{1}{|X|^{2m}}.
$$
\end{proof}

We are now ready to prove the analogue of Lemma~\ref{fewDiscounts} for
the efficient implementation of the median mechanism.
\begin{lemma}
\label{fewDiscountsEfficient}
For every set of $k$ queries $f_1,\ldots,f_k$, for all but an $O(|X|^{-m})$
fraction of fractional histograms $F$, the efficient implementation of
the median mechanism guarantees that:
The mechanism answers fewer than $40 m\log |X|$ hard queries,
except with probability $k\exp(-\Omega(\epsilon n\alpha'))$,
\end{lemma}

\begin{proof}
We assume that all answers to hard queries are $\epsilon/100$
accurate, and that $|r_i - \hat{r}_i| \leq
\frac{1}{100}$ for every~$i$.
By Lemmas~\ref{accurater} and~\ref{type2Accurate} --- the former adapted
to accommodate approximating $r_i$ via random sampling
--- we are in this case except with probability
$k\exp(-\Omega(\epsilon n\alpha'))$.

We analyze how the volume of
$C_i$ contracts with the number of hard queries answered.
Suppose the mechanism answers a hard query at time $i$. Then:
$$
r_i \leq \hat{r}_i + \frac{1}{100} < t_i +
\frac{1}{100} \leq \frac{91}{100}.
$$
Recall $C_i = \{F \in C_{i-1} : |f_i(F) - a_i| \leq \epsilon/50\}$.
Suppose that $\Vol(C_i) \geq \frac{94}{100}\Vol(C_{i-1})$. Then
$$
r_i =
\frac{\int_{C_{i-1}}\exp(-\eps^{-1}|f_i(F)-f_i(D)|)dF}{\Vol(C_{i-1})}$$$$
\geq \frac{94}{100}\exp\left(-\frac{1}{50}\right)
%\Vol(C_{i-1})
>
\frac{92}{100},
%\Vol(C_{i-1})
$$
a contradiction. Therefore, we have
\begin{equation}\label{eq:cts1}
|C_k| \leq \left(\frac{94}{100}\right)^h
 \Vol(C_{0}),
\end{equation}
if $h$ of the $k$ queries are hard.

Since all answers to hard queries are $\epsilon/100$ accurate, it must
be that $\textrm{Good}_{\epsilon/100}(D) \in C_k$.
Therefore, for an input database~$D$ that satisfies~\eqref{eq:volume}
--- and this is all but an $O(|X|^{-m})$ fraction of them, by
Lemma~\ref{goodVolumeLemma} --- we have
\begin{equation}\label{eq:cts2}
\Vol(C_k) \geq \Vol(\textrm{Good}_{\epsilon/100}(D)) \geq
\frac{\Vol(C_0)}{|X|^{2m}}.
\end{equation}
%where the last inequality holds except with negligible probability by
%lemma \ref{goodVolumeLemma}.
Combining inequalities~\eqref{eq:cts1} and~\eqref{eq:cts2} yields
$$
h \leq \frac{2m \ln |X|}{\ln \frac{50}{47}} < 40 m \ln |X|,
$$
as claimed.
\end{proof}

Lemmas~\ref{type2Accurate}, \ref{type1Accurate},
and~\ref{fewDiscountsEfficient} give the following utility guarantee.
\begin{theorem}
\label{efficientTheorem}
For every set $f_1,\ldots,f_k$ of queries, for all but a negligible
fraction of fractional histograms $F$, the efficient implementation of the
median mechanism is $(\epsilon,\delta)$-useful with
$\delta = k\exp(-\Omega(\epsilon n\alpha'))$.
\end{theorem}

\subsection{Usefulness for Finite Databases}
Fractional histograms correspond to probability distributions over $X$. Lemma \ref{goodVolumeLemma} shows that most probability distributions are `good' for the efficient implementation of the Median Mechanism; in fact, more is true. We next show that finite databases \emph{sampled} from randomly selected probability distributions also have good volume properties. Together, these lemmas show that the efficient implementation of the median mechanism will be able to answer nearly exponentially many queries with high probability, in the setting in which the private database $D$ is drawn from some `typical' population distribution.
\newline
\textbf{DatabaseSample}($|D|$):
\begin{enumerate}
\item Select a fractional point $F \in C_0$ uniformly at random.
\item Sample and return a database $D$ of size $|D|$ by drawing each $x \in D$ independently at random from the probability distribution over $X$ induced by $F$ (i.e. sample $x_i \in X$ with probability proportional to $F_i$).
\end{enumerate}
\begin{lemma}
\label{DatabaseSample}
For $|D|$ as in \eqref{eq:n} (as required for the Median Mechanism), a database sampled by \textbf{DatabaseSample}($|D|$) satisfies \eqref{eq:volume} except with probability at most $O(|X|^{-m})$.
\end{lemma}
\begin{proof}
By lemma \ref{goodVolumeLemma}, except with probability $|X|^{-m}$, the fractional histogram $F$ selected in step $1$ satisfies
$$\frac{\Vol(\textrm{Good}_{\epsilon/200}(F))}{\Vol(C_0)} \geq
\frac{1}{|X|^{2m}}.$$
By lemma \ref{uniformConvergence}, when we sample a database $D$ of size $|D| \geq O((\log |X|\log^3 k\log 1/\epsilon)/\epsilon^3)$ from the probability distribution induced by $F$, except with probability $\delta = O(k|X|^{-\log^3 k/\epsilon})$, $\textrm{Good}_{\epsilon/200}(F) \subset \textrm{Good}_{\epsilon/100}(D)$, which gives us condition \eqref{eq:volume}.
\end{proof}

\vspace{-0.1in}
We would like an analogue of lemma \ref{goodVolumeLemma} that holds for all but a diminishing fraction of \emph{finite} databases (which correspond to lattice points within $C_0$) rather than fractional points in $C_0$, but it is not clear how uniformly randomly sampled lattice points distribute themselves with respect to the volume of $C_0$. If $n >> |X|$, then the lattice will be fine enough to approximate the volume of $C_0$, and lemma \ref{goodVolumeLemma} will continue to hold. We now show that \emph{small} uniformly sampled databases will also be good for the efficient version of the median mechanism. Here, small means $n = o(\sqrt{|X|})$, which allows for databases which are still polynomial in the size of $X$. A tighter analysis is possible, but we opt instead to give a simple argument.

\vspace{-0.1in}
\begin{lemma}
\label{goodVolumeUniform}
For every $n$ such that $n$ satisfies \eqref{eq:n} and $n = o(\sqrt{|X|})$, all but an $O(n^2/|X|)$ fraction of databases $D$ of size $|D| = n$ satisfy condition \eqref{eq:volume}.
\end{lemma}
\begin{proof}
We proceed by showing that our \textbf{DatabaseSample} procedure, which we know via lemma \ref{DatabaseSample} generates databases that satisfy \eqref{eq:volume} with high probability, is close to uniform. Note that \textbf{DatabaseSample} first selects a \emph{probability distribution} $F$ uniformly at random from the positive quadrant of the $\ell_1$ ball, and then samples $D$ from $F$.

For any particular database $D^*$ with $|D^*| = n$ we write $\Pr_U[D = D^*]$ to denote the probability of generating $D^*$ when we sample a database uniformly at random, and we write $\Pr_N[D=D^*]$ to denote the probability of generating $D^*$ when we sample a database according to \textbf{DatabaseSample}. Let $R$ denote the event that $D^*$ contains no duplicate elements. We begin by noting by symmetry that:
$\Pr_U[D=D^*| R] = \Pr_N[D=D^*| R]$
We first argue that $\Pr_U[R]$ and $\Pr_N[R]$ are both large. We immediately have that the expected number of repetitions in database $D$ when drawn from the uniform distribution is ${n \choose 2}/|X|$, and so $\Pr_U[\neg R] \leq \frac{n^2}{|X|}$.
\sloppy
We now consider $\Pr_N[R]$. Since $F$ is a uniformly random point in the positive quadrant of the $\ell_1$ ball, each coordinate $F_i$ has the marginal of a Beta distribution: $F_i \sim \beta(1,|X|-1)$. (See, for example, \cite{Devroye} Chapter 5). Therefore, $E[F_i^2] = \frac{2}{|X|(|X|+1)}$ and so the expected number of repetitions in database $D$ when drawn from \textbf{DatabaseSample} is ${n \choose 2}\sum_{i=1}^{|X|}E[F_i^2]=\frac{2{n\choose 2}}{|X|+1}\leq \frac{2n^2}{|X|}$. Therefore, $\Pr_N[\neg R] \leq \frac{2n^2}{|X|}$.

Finally, let $B$ be the event that database $D$ fails to satisfy \eqref{eq:volume}. We have:
\begin{eqnarray*}
\Pr_U[B] &=& \Pr_U[B | R]\cdot\Pr_U[R] + \Pr_U[B|\neg R]\cdot\Pr_U[\neg R] \\
&=& \Pr_N[B | R]\cdot\Pr_U[R] + \Pr_U[B|\neg R]\cdot\Pr_U[\neg R] \\
&\leq& \Pr_N[B | R]\cdot\Pr_U[R] + \Pr_U[\neg R] \\
&\leq& \Pr_N[B]\cdot \frac{\Pr_U[R]}{\Pr_N[R]} + \Pr_U[\neg R] \\
&\leq& \frac{\Pr_N[B]}{1-\frac{2n^2}{|X|}} + \frac{n^2}{|X|} \\
&=& O(\frac{n^2}{|X|})
\end{eqnarray*}
where the last equality follows from lemma \ref{DatabaseSample}, which states that $\Pr_N[B]$ is negligibly small.
\end{proof}
We observe that we can substitute either of the above lemmas for lemma \ref{goodVolumeLemma} in the proof of lemma \ref{fewDiscountsEfficient} to obtain versions of Thoerem \ref{efficientTheorem}:
\begin{corollary}
For every set $f_1,\ldots,f_k$ of queries, for all but a negligible
fraction of databases sampled by \textbf{DatabaseSample}, the efficient implementation of the
median mechanism is $(\epsilon,\delta)$-useful with
$\delta = k\exp(-\Omega(\epsilon n\alpha'))$.
\end{corollary}
\begin{corollary}
For every set $f_1,\ldots,f_k$ of queries, for all but an $n^2/|X|$
fraction of uniformly randomly sampled databases of size $n$, the efficient implementation of the
median mechanism is $(\epsilon,\delta)$-useful with
$\delta = k\exp(-\Omega(\epsilon n\alpha'))$.
\end{corollary}
\section{Conclusion}
We have shown that in the setting of predicate queries, interactivity does not pose an information theoretic barrier to differentially private data release. In particular, our dependence on the number of queries $k$ nearly matches the optimal dependence of $\log k$ achieved in the \emph{offline} setting by \cite{BLR08}. We remark that our dependence on other parameters is not necessarily optimal: in particular, \cite{DNRRV09} achieves a better (and optimal) dependence on $\epsilon$. We have also shown how to implement our mechanism in time poly$(|X|,k)$, although at the cost of sacrificing worst-case utility guarantees. The question of an interactive mechanism with poly$(|X|,k)$ runtime and worst-case utility guarantees remains an interesting open question. More generally, although the lower bounds of \cite{DNRRV09} seem to preclude mechanisms with run-time poly$(\log |X|)$ from answering a superlinear number of generic predicate queries, the question of achieving this runtime for specific query classes of interest (offline or online) remains largely open. Recently a representation-dependent impossibility result for the class of conjunctions was obtained by Ullman and Vadhan \cite{UV10}: either extending this to a representation-independent impossibility result, or circumventing it by giving an efficient mechanism with a novel output representation would be very interesting.

\section*{Acknowledgments}

The first author wishes to thank a number of people for useful
discussions, including Avrim Blum, Moritz Hardt, Katrina Ligett, Frank
McSherry, and Adam Smith. He would particularly like to thank Moritz
Hardt for suggesting trying to prove usefulness guarantees for a
continuous version of the BLR mechanism, and Avrim Blum for suggesting
the distribution from which we select the threshold in the median
mechanism.

\bibliographystyle{alpha}
%\bibliography{OnlinePrivacy}
\newcommand{\etalchar}[1]{$^{#1}$}

\end{document}